\documentclass[sigconf]{acmart}

\usepackage{booktabs} 
\usepackage{color}
\usepackage{url}
\usepackage{multirow}
\usepackage{rotating}
\usepackage{array}
\usepackage{color}
\usepackage{verbatim}
\usepackage{bigstrut}
\usepackage{multirow}
\usepackage{amsmath}
\usepackage{amssymb}
\usepackage{setspace}
\usepackage{url}
\usepackage{pifont}
\usepackage{subfigure}
\usepackage{pifont}
\usepackage{afterpage}
\usepackage{rotating}
\usepackage{wrapfig}
\usepackage{caption}
\usepackage{graphicx}
\usepackage{caption}
\usepackage{enumitem}
\usepackage{xcolor}
\definecolor{pink}{rgb}{0.9,0,0.9}
\definecolor{gray}{rgb}{0.95,0.95,0.85}
\definecolor{mauve}{rgb}{0.1,0.7,0.2}
\definecolor{bg}{rgb}{0.95,0.95,0.9}
\definecolor{dkgreen}{rgb}{0,0.6,0}
\usepackage[utf8]{inputenc}
\usepackage{epstopdf}
\usepackage{tabulary}
\usepackage{float}
\usepackage{xspace}
\usepackage{flushend}

\newcolumntype{M}[1]{>{\centering\arraybackslash}m{#1}}
\usepackage{xcolor}
\definecolor{dkgreen}{rgb}{0,0.6,0}
\usepackage{listings}
\lstdefinelanguage{Lambda}{%
  morekeywords={%
    if,then,else,fix 
  },%
  morekeywords={[2]int},   
  otherkeywords={:}, 
  literate={
    {->}{{$\to$}}{2}
    {lambda}{{$\lambda$}}{1}
  },
  basicstyle={\sffamily},
  keywordstyle={\bfseries},
  keywordstyle={[2]\itshape}, 
  keepspaces,
  mathescape 
}[keywords,comments,strings]%

\setcopyright{acmcopyright}

\acmConference[ICSE'2018]{ICSE}{May 2018}{Gothenburg, Sweden} 
\acmYear{2018}
\copyrightyear{2018}

\PassOptionsToPackage{bookmarks=false}{hyperref}
\begin{document}

\title[Dronology Research Environment]{Dronology: An Incubator for Cyber-Physical Systems Research}

\author{Jane Cleland-Huang, Michael Vierhauser, Sean Bayley}
\affiliation{%
  \institution{Department of Computer Science and Engineering\\University of Notre Dame}
  \city{South Bend, IN} 
  \state{USA} 
}
\email{{JaneClelandHuang,mvierhau,sbayley}@nd.edu}

\renewcommand{\shortauthors}{Cleland-Huang et al.}

\begin{abstract}
Research in the area of Cyber-Physical Systems (CPS) is hampered by the lack of available project environments in which to explore open challenges and to propose and rigorously evaluate solutions. In this ``New Ideas and Emerging Results'' paper we introduce a CPS research incubator -- based upon a system, and its associated project environment, for managing and coordinating the flight of small Unmanned Aerial Systems (sUAS). The research incubator provides a new community resource, making available diverse, high-quality project artifacts produced across multiple releases of a safety-critical CPS. It enables researchers to experiment with their own novel solutions within a fully-executable runtime environment that supports both high-fidelity sUAS simulations as well as physical sUAS. Early collaborators from the software engineering community have shown broad and enthusiastic support for the project and its role as a research incubator, and have indicated their intention to leverage the environment to address their own research areas of goal modeling, runtime adaptation, safety-assurance, and software evolution.

\end{abstract}





\maketitle

\section{Introduction}
\label{sec:Intro}

The emerging adoption of Cyber-Physical Systems (CPS) in areas ranging across autonomous driving, smart cities, and unmanned aerial systems makes it imperative to address software engineering (SE) challenges for safety-critical systems~  \cite{DBLP:conf/dagstuhl/2011models,DBLP:conf/icse/NuseibehE00}. Challenges are diverse and include  goal modeling \cite{DBLP:journals/re/CailliauL13}, formal methods, requirements completeness \cite{DBLP:conf/models/DeVriesC16}, runtime adaptation, \cite{lotus}, design-time evolution~\cite{DBLP:journals/ase/DevineGKL16}, product line development~\cite{Kruger2017SPLC}, and human-computer interaction, to name a few.  However, such research cannot be conducted in a vacuum, and must be motivated, explored, and evaluated within the context of realistic and real-world systems. 

The Dronology\ system, and its associated project environment serves as a \emph{research incubator} to facilitate and nurture early growth of research ideas in the CPS domain. It enables researchers to experiment with a theory or hypothesis, in a controlled environment, and to progressively develop the idea until it is ready for testing and deployment in a full industrial setting. Currently many software engineering concepts do not progress past inception because researchers lack the means to advance them through intermediate stages of growth.  Dronology provides a full \emph{project environment} (i.e., a fully functional open-source system) for managing, monitoring, and coordinating the flights of multiple small Unmanned Aerial Systems (sUAS) and a \emph{dataset} including diverse, high-quality project artifacts produced across multiple releases.  As such, it provides a realistic incubator for investigating CPS topics, and for enabling researchers to experiment with their own innovative solutions within a robust project environment. Furthermore, as sUAS accidents, such as in-air collisions, or crash landings into vehicles or pedestrians, have the potential for causing significant harm to people and/or property damage. Dronology is therefore considered \emph{safety critical}.

To illustrate the need for a CPS research incubator, consider a researcher working in the space of trace link evolution and reuse in order to support safety analysis in a safety-critical product line. His work is greatly impeded by the lack of publicly available, non-trivially sized, multi-version, software systems containing the type of diverse software artifacts and feature definitions that are expected in a safety-critical product line.  Similarly, a researcher working in the space of CPS runtime adaptation has developed her own experimental environment for proof-of-concept exploration; however, she is interested in evaluating her work on runtime composition of features in a larger-scale, real-world environment. The current lack of such a publicly available environment is a significant impediment to research in the CPS space. We aim to address this issue through our proposed CPS research incubator.  

In the remainder of this paper, Section \ref{sec:CaseEnvs} discusses shortcomings of existing community resources. Sections~\ref{sec:researchEnvironment} and \ref{sec:researchAreas}, describe the Dronology project environment and provide examples of research areas that the incubator is designed to support. Finally, in Section \ref{sec:Conclusion} we describe ongoing goals for our proposed research incubator.

\section{Existing Research Environments}
\label{sec:CaseEnvs}

One of the most closely related research contributions is the PROMISE repository established by Menzies et al. \cite{promiserepo}.  PROMISE includes an extensive collection of diverse software engineering artifacts -- contributed by numerous SE researchers and used for countless experiments. However, a major limitation of the repository with respect to CPS research stems from the fact that the hosted datasets tend to represent simple snapshots of specific parts of a system and neither provide executable code nor diverse project artifacts. Recent history has taught us that researchers actively pursue open challenges that are supported by appropriate datasets.  A compelling example is the concurrent growth of open-source system (OSS) projects with increased research emphasis on \emph{mining software repositories} (MSR).  Our goal is to empower CPS researchers in the same way that OSS empowered MSR researchers. In the remainder of this section we summarize the reasons that current datasets and environments are insufficient to meet CPS research needs. \newline\vspace{-8pt}

\noindent{\bf Open Source Systems:} 
OSS projects provide a rich source of data for many areas of investigation. However they are  more suited to research that focuses primarily on source code and related artifacts, such as bug reports, test cases, or developer interactions, and do not provide suitable environments for studying other types of challenges, such as the tight interaction between hardware and software that is prevalent in safety-critical and/or Cyber-Physical Systems.\newline\vspace{-8pt} 

\noindent{\bf Existing Community Datasets:}
Several community datasets support focused research efforts.  For example the Generic Infusion Pump (GIP), includes requirements, architectural designs, and formal models and has been used in many studies~\cite{DBLP:conf/sigada/MurugesanWRH13}. However, its artifacts form a relatively disjoint set and do not include source code. 
The iTrust system includes use cases, source code, and limited (or outdated) trace links \cite{iTrust} but represents  an IT system and is therefore also not suited to CPS research. Similarly, datasets provided by PROMISE \cite{promiserepo}, the Center of Excellence for Software Traceability \cite{DBLP:conf/re/Cleland-HuangCH13}, and the SIRS testing repository \cite{Do:2005:SCE:1089922.1089928}, are widely used by software engineering researchers but only provide data snapshots and lack complete and comprehensive project environments. \newline\vspace{-8pt}

\noindent{\bf Industry Datasets: }
Industrial datasets provide rich research environments. However, finding industry collaborators can be challenging, and data that is shared under non-disclosure agreements can not be shared openly, meaning that researchers can not compare their solutions or reveal meaningful contextual details. Therefore, a pressing need exists for a publicly accessible safety-critical CPS project that provides a rich project context for enabling experimentation across the diverse challenges of CPSs. 

\begin{figure}[!t]
    \centering
    \includegraphics[width=1.0\columnwidth]{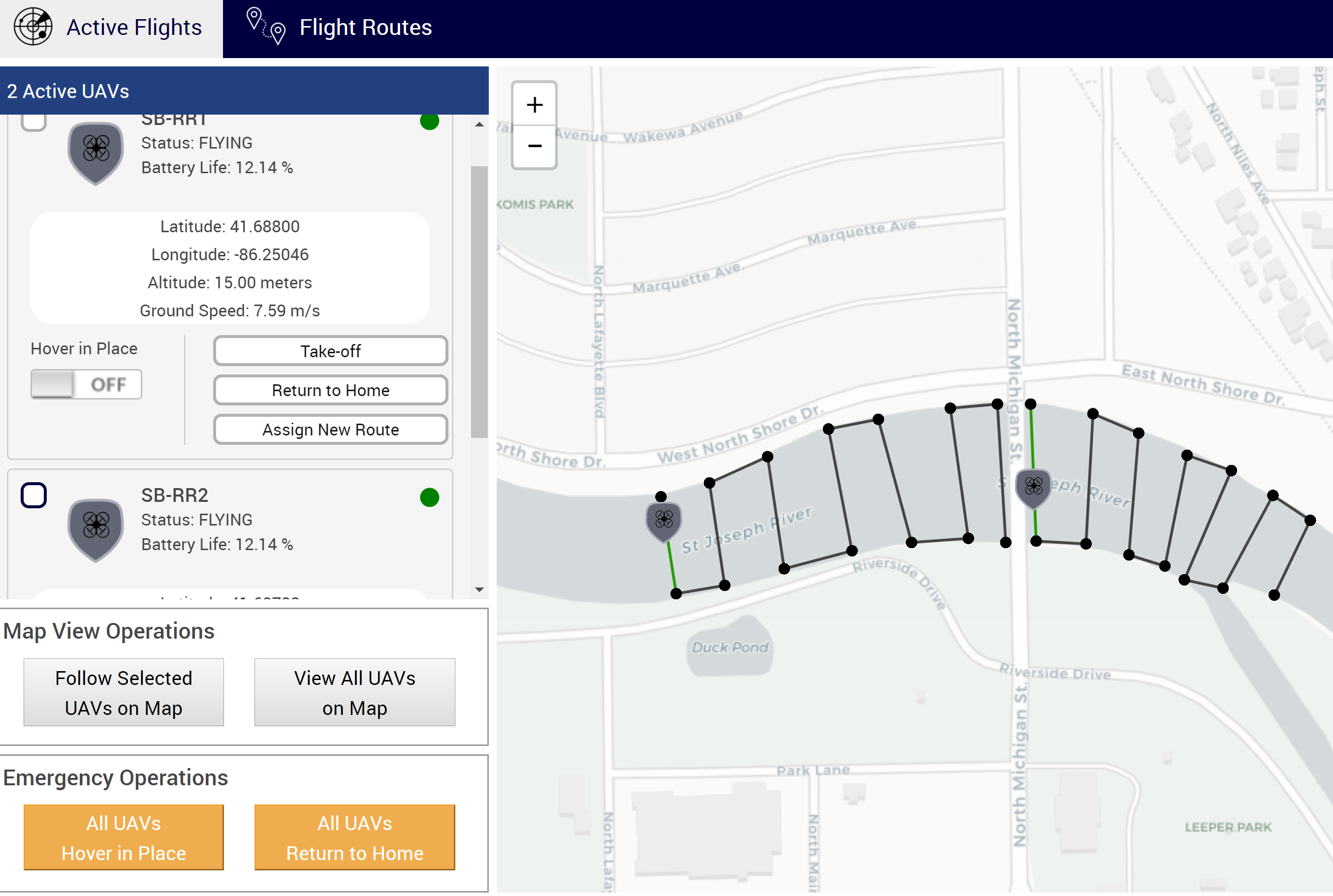}
    \caption{Dronology: sUAS flight routes displayed and monitored in real-time}
    \label{fig:xDronesUI}
    \vspace{-8pt}
\end{figure}

\section{Dronology: A Research Incubator}
\label{sec:researchEnvironment}
We established criteria for the CPS research incubator that it must (1) include a cyber-physical component, (2) use affordable and readily available hardware, (3) be accessible to a broad set of researchers in terms of  the necessary domain knowledge, (4) include diverse and challenging safety hazards, and (5) be developed using standards commensurate with medium criticality projects.

\subsection{Dronology Product}
Dronology is developed as a Java-based framework for controlling, managing, monitoring, and coordinating flights of sUAS in an urban airspace.  It is compatible with ArduPilot-based\footnote{\url{http://ardupilot.org}} sUAS and integrates a high-fidelity Software-in-the-loop (SITL) simulator as well as physical sUAS via a Python-based groundstation. 

To develop a realistic, industrial-strength CPS environment we are working with end-users on two concrete projects to support river rescue and defibrillator delivery. Figure \ref{fig:xDronesUI} illustrates the river-rescue scenario in our Dronology UI. We are also working closely with researchers from four diverse research groups to ensure that the Dronology architecture, process, and artifacts actually support researcher's needs in their areas of interest.

\begin{figure}[!t]
    \centering
    \includegraphics[width=.9\columnwidth]{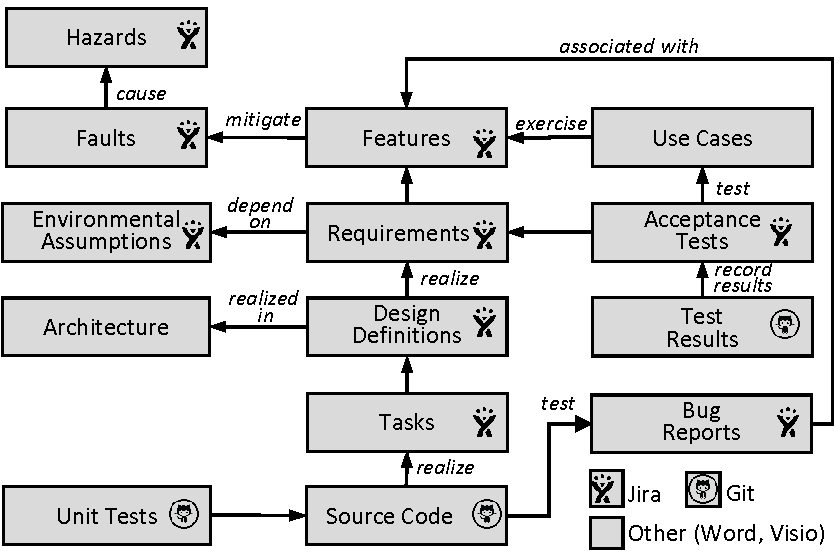}
    \caption{Dronology' Artifacts and Traceability Paths}
    \label{fig:artifacts}
\end{figure}

\begin{figure}[!t]
    \centering
    \includegraphics[width=1.0\columnwidth]{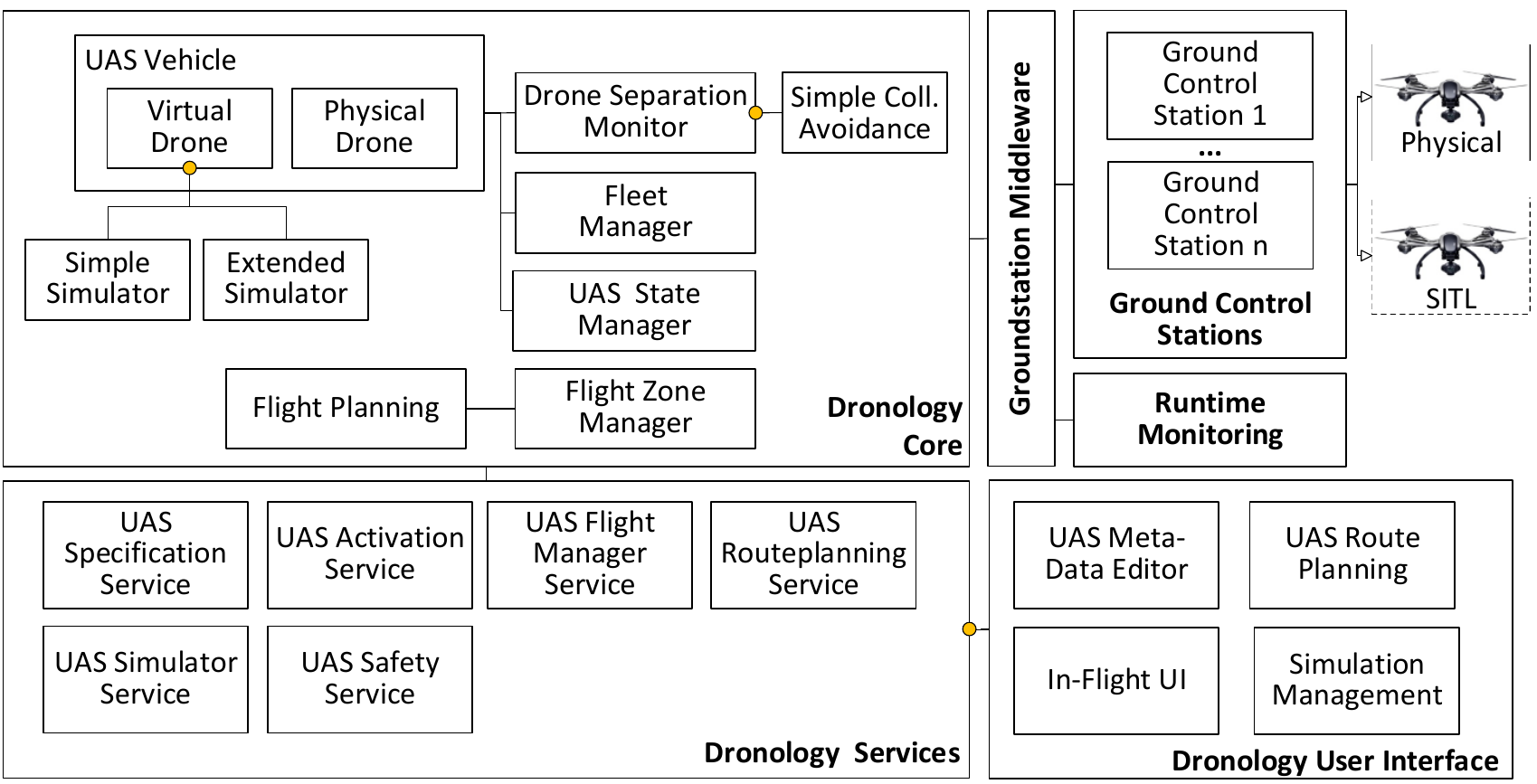}
    \caption{High level Architecture of Dronology}
    \label{fig:architecture}
\end{figure}

The project produces diverse artifacts such as hazards, faults, environmental assumptions, requirements, design definitions, architectural design, a feature model, use cases, acceptance tests with test logs, unit tests, source code, and bug reports as shown in Figure \ref{fig:artifacts}.  Though incomplete, this set of artifacts is extensible and represents fundamental elements of  safety-critical systems~\cite{DBLP:conf/se/RempelMKC14}. 
Version 0.1, which will be released in Spring of 2018, currently includes approximately 20,000 lines of
code (in around 300 Java classes and 30 Python classes)
 a feature model containing 39 features, around 250 requirements alongside with 150 design definitions, 95 bug reports, numerous safety-related artifacts, and the architecture shown in Figure \ref{fig:architecture}.  A current description of the project goals, its current status, sample artifacts, process description, and release timeline is available at \url{http://www.dronology.info}.

The project environment will be available upon request via github, where researchers can download or fork the project to build their own new features, extensions, and modifications. In the future it will be fully open-sourced. As it is common in many open-source projects we will establish a  review process to integrate contributions of newly developed features that warrant inclusion in future official releases. 

\subsection{Dronology Process}
Dronology seeks to deliver a high-quality, industrial-like project environment.  To this end, our initial project team includes a hired professional developer with previous experience working on sUAS applications, four team members with significant industrial programming experience, a hardware expert who is a qualified sUAS pilot, one team member with extensive experience working in safety-critical systems, in addition to several graduate and undergraduate students who have completed internships on the project. Finally, we are establishing an advisory board of actively engaged industrial experts, some of whom have already provided feedback on the project.

The difference between our system and industrial systems is therefore not in the scope, completeness, or quality of the delivered system, but rather in the goals that drive its development. Our goals are two-fold.  First, to \emph{build a safe, deployed, working system} in conjunction with real end-users, and second, to \emph{create and maintain a research incubator that meets the needs} of researchers working in diverse areas of safety-critical systems and cyber-physical systems.

\subsection{Limitations}
While we are convinced that our research incubator supports the needs of researchers in diverse areas, there are certain limitations that need to be taken into consideration.
For example, using Java and Python as our main implementation languages might limit the suitability for certain areas of real-time critical CPS research, meaning that in its current form Dronology does not represent a system which exhibits hard real-time characteristics. While there have been efforts to use Java in real time embedded systems~\cite{Sharp2003RealTimeJava} this is out of scope of the (core) application area of the research incubator. However, the product-line approach allows us to add additional features or alternative implementations later as needed.

\section{Supported Research Areas}
\label{sec:researchAreas}
Research challenges associated with CPS and other types of safety-critical system are diverse.  We present six topic areas and briefly describe aspects of the Dronology incubator that would support research in each area.  This list is not intended to be complete. \newline\vspace{-8pt}

\noindent\textbf{$\bullet$ Software and Systems Requirements: } In their roadmap paper, Nuseibeh and Easterbrook~\cite{DBLP:conf/icse/NuseibehE00} outlined specific challenges associated with eliciting, specifying, analyzing, modeling, and using requirements.  Almost all of these challenges are relevant in the context of CPS.  Furthermore, several specific problems, such as requirements analysis and modeling, are particularly impacted by the CPS nature of the project and should be explored within such an environment.  The Dronology environment creates opportunities for research in several of these areas, for example, creating requirements models that represent data, domain, and behavioral views of the system.  Dronology includes goals, system and software level requirements, and design definitions that provide necessary foundations for requirements analysis and modeling.\newline \vspace{-8pt}

\noindent\textbf{$\bullet$ Software Traceability:}
Software traceability captures relationships between uniquely identifiable software engineering artifacts and is required for certification (e.g., DO-178C) in most safety-critical domains. Specific challenges laid out in a traceability roadmap on ``trends and future directions'' include trace strategizing, trace link creation and evolution, and trace link usage and visualization~ \cite{DBLP:conf/icse/Cleland-HuangGHMZ14}.  The diversity of artifacts in the Dronology project environment, coupled with its extensive network of trace links, and the project's full history of versions and commits enables broad research across all open research areas. Such extensive and diverse sets of artifacts are  not available in any datasets currently provided by the Center of Excellence for Software Traceability \cite{DBLP:conf/re/Cleland-HuangCH13} or in any other dataset we are aware of. \newline\vspace{-8pt}

\noindent\textbf{$\bullet$ CPS Product Lines:} Many CPSs could appropriately be developed as product lines in which a set of products that share a common, managed set of features  are developed from a common set of core assets  in a prescribed way \cite{Clements2001}. Researchers in this area address diverse topics at both the domain and application engineering levels \cite{Kruger2017SPLC} including variability modeling and analysis, consistency between applications and domain artifacts~\cite{vierhauserCC}, formal verification and testing, as well as application realization, feature interactions~\cite{Apel2011FA}, dynamic product generation, and safety assurance of individual products. Datasets provided within the Product Line (PL) community tend to focus on feature models and do not provide complete project environments. Dronology is developed as a CPS product line with a respective feature model, PL architecture, variability points, and a configuration mechanism, and therefore provides an environment for PL experimentation.\newline\vspace{-8pt}

\noindent\textbf{$\bullet$ Safety Assurance:}
Most CPSs are safety-critical -- leading to research in areas of formal modeling and verification, testing, simulation, hazard analysis, and safety assurance \cite{Leveson:1995:SSS:202709}. In this space, systems can be developed both formally and informally -- although both approaches require a rigorous, safety-imbued engineering process. The Dronology process includes a preliminary hazard analysis, a failure model effect criticality analysis (FMECA), mitigating requirements that address functional concerns and architectural solutions such as fault-tolerance, performance, and reliability, and Safety Assurance Cases~\cite{Kelly2004SAC}. We plan a limited set of formal models (e.g., in the areas of collision avoidance and obstacle detection); however researchers could build their own models based on Dronology requirements, design definitions, and associated properties. \newline\vspace{-8pt}

\noindent\textbf{$\bullet$ Runtime Monitoring and Adaption:} CPSs often need to adapt at runtime to changes in their environments. This introduces a complex and multidisciplinary research agenda \cite{deLemos2013}, that includes modeling and monitoring~\cite{lotus} causes of adaptation, specifying and constructing mechanisms of adaptation, and analyzing effects of adaptation upon the system and its safety~\cite{Calinescu2012SelfAdaptive}. 
Dronology provides an extensible runtime monitoring infrastructure allowing researchers to monitor diverse data collected from UAV sensors as well as internally generated events such as flight mode transitions. Adaptation strategies can be evaluated at variability points in the product line or through modifying cloned source code. 
\newline\vspace{-8pt}

\noindent\textbf{$\bullet$ Human Studies:} Last, but not least, is the opportunity for supporting Human Studies within the context of safety-critical CPS. Such studies are also limited by a lack of immersive environments.  Example research topics include studying visualization techniques during  change impact analysis  \cite{DBLP:journals/tse/VaraBWM16}, studying the role of humans in trace link creation and maintenance \cite{DBLP:books/daglib/p/DekhtyarH12}, and studying ways in which humans leverage tools to analyze system safety.  The Dronology project environment creates a viable context for such studies.

\section{Future Vision}
\label{sec:Conclusion}
We have introduced our vision for Dronology as a research incubator for safety-critical CPS. Future activities include: (1) \emph{Ongoing development} of numerous features to be delivered across multiple versions; (2) A  \emph{product line} with multiple variability points and assets to be delivered at both the domain and application level;
 (3) \emph{Research resources } to facilitate research challenges identified by the community, by organizing artifacts as customized downloadable data bundles; (4) An  \emph{atypical OSS } for which contributors will provide not just source code, but other safety-related artifacts required by the Dronology project;  (5) \emph{ real end users } to bring realism to the project; (6) \emph{Incremental releases } starting in May, 2018.

Our research incubator goes far beyond a more traditional dataset consisting of snapshots of data. It provides a rich project environment that includes diverse artifacts commensurate with a safety-critical domain.  Our incubator is designed to empower researchers working in areas of CPS research which suffer from a dearth of experimental project environments.  Dronology is planned as a community endeavor and collaborators are welcomed to work with us to make this a long-term, effective community resource. Project artifacts may be downloaded directly from \url{http://www.dronology.info} while source code is available upon request by research groups.

\section*{Acknowledgments}

This project has been funded by the Austrian Science Fund~(FWF) (J3998-N319) and US National Science Foundation Grants (CCF-1741781, CCF-1649448)

\bibliographystyle{ACM-Reference-Format}
\bibliography{References} 


\begin{thebibliography}{00}


\ifx \showCODEN    \undefined \def \showCODEN     #1{\unskip}     \fi
\ifx \showDOI      \undefined \def \showDOI       #1{#1}\fi
\ifx \showISBNx    \undefined \def \showISBNx     #1{\unskip}     \fi
\ifx \showISBNxiii \undefined \def \showISBNxiii  #1{\unskip}     \fi
\ifx \showISSN     \undefined \def \showISSN      #1{\unskip}     \fi
\ifx \showLCCN     \undefined \def \showLCCN      #1{\unskip}     \fi
\ifx \shownote     \undefined \def \shownote      #1{#1}          \fi
\ifx \showarticletitle \undefined \def \showarticletitle #1{#1}   \fi
\ifx \showURL      \undefined \def \showURL       {\relax}        \fi
\providecommand\bibfield[2]{#2}
\providecommand\bibinfo[2]{#2}
\providecommand\natexlab[1]{#1}
\providecommand\showeprint[2][]{arXiv:#2}

\bibitem[\protect\citeauthoryear{Apel, Speidel, Wendler, von Rhein, and
  Beyer}{Apel et~al\mbox{.}}{2011}]%
        {Apel2011FA}
\bibfield{author}{\bibinfo{person}{Sven Apel}, \bibinfo{person}{Hendrik
  Speidel}, \bibinfo{person}{Philipp Wendler}, \bibinfo{person}{Alexander von
  Rhein}, {and} \bibinfo{person}{Dirk Beyer}.} \bibinfo{year}{2011}\natexlab{}.
\newblock \showarticletitle{Detection of Feature Interactions Using
  Feature-aware Verification}. In \bibinfo{booktitle}{{\em Proc. of the 2011
  26th IEEE/ACM Int'l Conf. on Automated Software Engineering}}.
  \bibinfo{publisher}{IEEE}, \bibinfo{pages}{372--375}.
\newblock


\bibitem[\protect\citeauthoryear{Barbosa, Lima, Maia, and Costa}{Barbosa
  et~al\mbox{.}}{2017}]%
        {lotus}
\bibfield{author}{\bibinfo{person}{Davi~Monteiro Barbosa},
  \bibinfo{person}{Romulo Gadelha De~Moura Lima}, \bibinfo{person}{Paulo
  Henrique~Mendes Maia}, {and} \bibinfo{person}{Evilasio Costa}.}
  \bibinfo{year}{2017}\natexlab{}.
\newblock \showarticletitle{Lotus@Runtime: A Tool for Runtime Monitoring and
  Verification of Self-Adaptive Systems}. In \bibinfo{booktitle}{{\em Proc. of
  the 12th Int{\textquotesingle}l Symp. on Software Engineering for Adaptive
  and Self-Managing Systems}}. \bibinfo{publisher}{IEEE}.
\newblock


\bibitem[\protect\citeauthoryear{Bencomo, France, Cheng, and
  A{\ss}mann}{Bencomo et~al\mbox{.}}{2014}]%
        {DBLP:conf/dagstuhl/2011models}
\bibfield{editor}{\bibinfo{person}{Nelly Bencomo}, \bibinfo{person}{Robert~B.
  France}, \bibinfo{person}{Betty H.~C. Cheng}, {and} \bibinfo{person}{Uwe
  A{\ss}mann}} (Eds.). \bibinfo{year}{2014}\natexlab{}.
\newblock \bibinfo{booktitle}{{\em Models@run.time - Foundations, Applications,
  and Roadmaps [Dagstuhl Seminar 11481, 2011]}}. \bibinfo{series}{Lecture Notes
  in Computer Science}, Vol.~\bibinfo{volume}{8378}.
  \bibinfo{publisher}{Springer}.
\newblock
\showISBNx{978-3-319-08914-0}


\bibitem[\protect\citeauthoryear{Cailliau and van Lamsweerde}{Cailliau and van
  Lamsweerde}{2013}]%
        {DBLP:journals/re/CailliauL13}
\bibfield{author}{\bibinfo{person}{Antoine Cailliau} {and}
  \bibinfo{person}{Axel van Lamsweerde}.} \bibinfo{year}{2013}\natexlab{}.
\newblock \showarticletitle{Assessing requirements-related risks through
  probabilistic goals and obstacles}.
\newblock \bibinfo{journal}{{\em Requir. Eng.\/}} \bibinfo{volume}{18},
  \bibinfo{number}{2} (\bibinfo{year}{2013}), \bibinfo{pages}{129--146}.
\newblock


\bibitem[\protect\citeauthoryear{Calinescu, Ghezzi, Kwiatkowska, and
  Mirandola}{Calinescu et~al\mbox{.}}{2012}]%
        {Calinescu2012SelfAdaptive}
\bibfield{author}{\bibinfo{person}{Radu Calinescu}, \bibinfo{person}{Carlo
  Ghezzi}, \bibinfo{person}{Marta Kwiatkowska}, {and} \bibinfo{person}{Raffaela
  Mirandola}.} \bibinfo{year}{2012}\natexlab{}.
\newblock \showarticletitle{Self-adaptive software needs quantitative
  verification at runtime}.
\newblock \bibinfo{journal}{{\it Commun. ACM}} \bibinfo{volume}{55},
  \bibinfo{number}{9} (\bibinfo{year}{2012}), \bibinfo{pages}{69--77}.
\newblock


\bibitem[\protect\citeauthoryear{Cleland{-}Huang, Czauderna, and
  Hayes}{Cleland{-}Huang et~al\mbox{.}}{2013}]%
        {DBLP:conf/re/Cleland-HuangCH13}
\bibfield{author}{\bibinfo{person}{Jane Cleland{-}Huang}, \bibinfo{person}{Adam
  Czauderna}, {and} \bibinfo{person}{Jane~Huffman Hayes}.}
  \bibinfo{year}{2013}\natexlab{}.
\newblock \showarticletitle{Using tracelab to design, execute, and baseline
  empirical requirements engineering experiments}. In \bibinfo{booktitle}{{\em
  Proc. of the 21st {IEEE} Intn'l Requirements Eng. Conf.}}
  \bibinfo{pages}{338--339}.
\newblock


\bibitem[\protect\citeauthoryear{Cleland{-}Huang, Gotel, Hayes, M{\"{a}}der,
  and Zisman}{Cleland{-}Huang et~al\mbox{.}}{2014}]%
        {DBLP:conf/icse/Cleland-HuangGHMZ14}
\bibfield{author}{\bibinfo{person}{Jane Cleland{-}Huang},
  \bibinfo{person}{Orlena Gotel}, \bibinfo{person}{Jane~Huffman Hayes},
  \bibinfo{person}{Patrick M{\"{a}}der}, {and} \bibinfo{person}{Andrea
  Zisman}.} \bibinfo{year}{2014}\natexlab{}.
\newblock \showarticletitle{Software traceability: trends and future
  directions}. In \bibinfo{booktitle}{{\em Proc. of the on Future of Software
  Engineering}}. \bibinfo{pages}{55--69}.
\newblock


\bibitem[\protect\citeauthoryear{Clements and Northrop}{Clements and
  Northrop}{2001}]%
        {Clements2001}
\bibfield{author}{\bibinfo{person}{Paul~C. Clements} {and}
  \bibinfo{person}{Linda Northrop}.} \bibinfo{year}{2001}\natexlab{}.
\newblock \bibinfo{booktitle}{{\em Software Product Lines: Practices and
  Patterns}}.
\newblock \bibinfo{publisher}{Addison-Wesley}.
\newblock


\bibitem[\protect\citeauthoryear{de~la Vara, Borg, Wnuk, and Moonen}{de~la Vara
  et~al\mbox{.}}{2016}]%
        {DBLP:journals/tse/VaraBWM16}
\bibfield{author}{\bibinfo{person}{Jose~Luis de~la Vara},
  \bibinfo{person}{Markus Borg}, \bibinfo{person}{Krzysztof Wnuk}, {and}
  \bibinfo{person}{Leon Moonen}.} \bibinfo{year}{2016}\natexlab{}.
\newblock \showarticletitle{An Industrial Survey of Safety Evidence Change
  Impact Analysis Practice}.
\newblock \bibinfo{journal}{{\em {IEEE} Trans. Software Eng.\/}}
  \bibinfo{volume}{42}, \bibinfo{number}{12} (\bibinfo{year}{2016}),
  \bibinfo{pages}{1095--1117}.
\newblock


\bibitem[\protect\citeauthoryear{De~Lemos, Giese, M{\"u}ller, Shaw, Andersson,
  Litoiu, Schmerl, Tamura, Villegas, Vogel, et~al\mbox{.}}{De~Lemos
  et~al\mbox{.}}{2013}]%
        {deLemos2013}
\bibfield{author}{\bibinfo{person}{Rog{\'e}rio De~Lemos},
  \bibinfo{person}{Holger Giese}, \bibinfo{person}{Hausi~A M{\"u}ller},
  \bibinfo{person}{Mary Shaw}, \bibinfo{person}{Jesper Andersson},
  \bibinfo{person}{Marin Litoiu}, \bibinfo{person}{Bradley Schmerl},
  \bibinfo{person}{Gabriel Tamura}, \bibinfo{person}{Norha~M Villegas},
  \bibinfo{person}{Thomas Vogel}, {et~al\mbox{.}}}
  \bibinfo{year}{2013}\natexlab{}.
\newblock \showarticletitle{Software engineering for self-adaptive systems: A
  second research roadmap}.
\newblock In \bibinfo{booktitle}{{\em Software Engineering for Self-Adaptive
  Systems II}}. \bibinfo{publisher}{Springer}, \bibinfo{pages}{1--32}.
\newblock


\bibitem[\protect\citeauthoryear{Dekhtyar and Hayes}{Dekhtyar and
  Hayes}{2012}]%
        {DBLP:books/daglib/p/DekhtyarH12}
\bibfield{author}{\bibinfo{person}{Alex Dekhtyar} {and}
  \bibinfo{person}{Jane~Huffman Hayes}.} \bibinfo{year}{2012}\natexlab{}.
\newblock \showarticletitle{Studying the Role of Humans in the Traceability
  Loop}.
\newblock In \bibinfo{booktitle}{{\em Software and Systems Traceability.}}
  \bibinfo{pages}{241--261}.
\newblock


\bibitem[\protect\citeauthoryear{Devine, Goseva{-}Popstojanova, Krishnan, and
  Lutz}{Devine et~al\mbox{.}}{2016}]%
        {DBLP:journals/ase/DevineGKL16}
\bibfield{author}{\bibinfo{person}{Thomas~R. Devine}, \bibinfo{person}{Katerina
  Goseva{-}Popstojanova}, \bibinfo{person}{Sandeep Krishnan}, {and}
  \bibinfo{person}{Robyn~R. Lutz}.} \bibinfo{year}{2016}\natexlab{}.
\newblock \showarticletitle{Assessment and cross-product prediction of software
  product line quality: accounting for reuse across products, over multiple
  releases}.
\newblock \bibinfo{journal}{{\em Autom. Softw. Eng.\/}} \bibinfo{volume}{23},
  \bibinfo{number}{2} (\bibinfo{year}{2016}), \bibinfo{pages}{253--302}.
\newblock


\bibitem[\protect\citeauthoryear{DeVries and Cheng}{DeVries and Cheng}{2016}]%
        {DBLP:conf/models/DeVriesC16}
\bibfield{author}{\bibinfo{person}{Byron DeVries} {and} \bibinfo{person}{Betty
  H.~C. Cheng}.} \bibinfo{year}{2016}\natexlab{}.
\newblock \showarticletitle{Automatic detection of incomplete requirements via
  symbolic analysis}. In \bibinfo{booktitle}{{\em Proc. of the {ACM/IEEE} 19th
  Int'l Conf. on Model Driven Engineering Languages and Systems}}.
  \bibinfo{pages}{385--395}.
\newblock


\bibitem[\protect\citeauthoryear{Do, Elbaum, and Rothermel}{Do
  et~al\mbox{.}}{2005}]%
        {Do:2005:SCE:1089922.1089928}
\bibfield{author}{\bibinfo{person}{Hyunsook Do}, \bibinfo{person}{Sebastian
  Elbaum}, {and} \bibinfo{person}{Gregg Rothermel}.}
  \bibinfo{year}{2005}\natexlab{}.
\newblock \showarticletitle{Supporting Controlled Experimentation with Testing
  Techniques: An Infrastructure and Its Potential Impact}.
\newblock \bibinfo{journal}{{\em Empirical Softw. Eng.\/}}
  \bibinfo{volume}{10}, \bibinfo{number}{4} (\bibinfo{date}{Oct.}
  \bibinfo{year}{2005}), \bibinfo{pages}{405--435}.
\newblock
\showISSN{1382-3256}


\bibitem[\protect\citeauthoryear{Kelly and Weaver}{Kelly and Weaver}{2004}]%
        {Kelly2004SAC}
\bibfield{author}{\bibinfo{person}{Tim Kelly} {and} \bibinfo{person}{Rob
  Weaver}.} \bibinfo{year}{2004}\natexlab{}.
\newblock \showarticletitle{The Goal Structuring Notation--a safety argument
  notation}. In \bibinfo{booktitle}{{\em WS on Assurance Cases of Dependable
  Systems and Networks}}.
\newblock


\bibitem[\protect\citeauthoryear{Kr\"{u}ger, Nielebock, Krieter, Diedrich,
  Leich, Saake, Zug, and Ortmeier}{Kr\"{u}ger et~al\mbox{.}}{2017}]%
        {Kruger2017SPLC}
\bibfield{author}{\bibinfo{person}{Jacob Kr\"{u}ger},
  \bibinfo{person}{Sebastian Nielebock}, \bibinfo{person}{Sebastian Krieter},
  \bibinfo{person}{Christian Diedrich}, \bibinfo{person}{Thomas Leich},
  \bibinfo{person}{Gunter Saake}, \bibinfo{person}{Sebastian Zug}, {and}
  \bibinfo{person}{Frank Ortmeier}.} \bibinfo{year}{2017}\natexlab{}.
\newblock \showarticletitle{Beyond Software Product Lines: Variability Modeling
  in Cyber-Physical Systems}. In \bibinfo{booktitle}{{\em Proc. of the 21st
  Int'l Systems and Software Product Line Conf.}} \bibinfo{publisher}{ACM},
  \bibinfo{pages}{237--241}.
\newblock


\bibitem[\protect\citeauthoryear{Leveson}{Leveson}{1995}]%
        {Leveson:1995:SSS:202709}
\bibfield{author}{\bibinfo{person}{Nancy~G. Leveson}.}
  \bibinfo{year}{1995}\natexlab{}.
\newblock \bibinfo{booktitle}{{\em Safeware: System Safety and Computers}}.
\newblock \bibinfo{publisher}{ACM}, \bibinfo{address}{New York, NY, USA}.
\newblock
\showISBNx{0-201-11972-2}


\bibitem[\protect\citeauthoryear{Menzies, Krishna, and Pryor}{Menzies
  et~al\mbox{.}}{2016}]%
        {promiserepo}
\bibfield{author}{\bibinfo{person}{Tim Menzies}, \bibinfo{person}{R. Krishna},
  {and} \bibinfo{person}{D. Pryor}.} \bibinfo{year}{2016}\natexlab{}.
\newblock \bibinfo{title}{The Promise Repository of Empirical Software
  Engineering Data"}.
\newblock   (\bibinfo{year}{2016}).
\newblock


\bibitem[\protect\citeauthoryear{Murugesan, Whalen, Rayadurgam, and
  Heimdahl}{Murugesan et~al\mbox{.}}{2013}]%
        {DBLP:conf/sigada/MurugesanWRH13}
\bibfield{author}{\bibinfo{person}{Anitha Murugesan},
  \bibinfo{person}{Michael~W. Whalen}, \bibinfo{person}{Sanjai Rayadurgam},
  {and} \bibinfo{person}{Mats Per~Erik Heimdahl}.}
  \bibinfo{year}{2013}\natexlab{}.
\newblock \showarticletitle{Compositional verification of a medical device
  system}. In \bibinfo{booktitle}{{\em ACM SIGAda Annual Conf. on High
  integrity language technology}}. \bibinfo{pages}{51--64}.
\newblock


\bibitem[\protect\citeauthoryear{Nuseibeh and Easterbrook}{Nuseibeh and
  Easterbrook}{2000}]%
        {DBLP:conf/icse/NuseibehE00}
\bibfield{author}{\bibinfo{person}{Bashar Nuseibeh} {and}
  \bibinfo{person}{Steve~M. Easterbrook}.} \bibinfo{year}{2000}\natexlab{}.
\newblock \showarticletitle{Requirements Engineering: a Roadmap}. In
  \bibinfo{booktitle}{{\em Proc. of the 22nd Int'l Conf. on on Software
  Engineering, Future of Software Engineering Track}}. \bibinfo{pages}{35--46}.
\newblock


\bibitem[\protect\citeauthoryear{Rempel, M{\"{a}}der, Kuschke, and
  Cleland{-}Huang}{Rempel et~al\mbox{.}}{2014}]%
        {DBLP:conf/se/RempelMKC14}
\bibfield{author}{\bibinfo{person}{Patrick Rempel}, \bibinfo{person}{Patrick
  M{\"{a}}der}, \bibinfo{person}{Tobias Kuschke}, {and} \bibinfo{person}{Jane
  Cleland{-}Huang}.} \bibinfo{year}{2014}\natexlab{}.
\newblock \showarticletitle{Mind the gap: assessing the conformance of software
  traceability to relevant guidelines}. In \bibinfo{booktitle}{{\em Proc. of
  the Int'l Conf. on Software Engineering}}. \bibinfo{pages}{943--954}.
\newblock


\bibitem[\protect\citeauthoryear{Sharp, Pla, Luecke, and Hassan}{Sharp
  et~al\mbox{.}}{2003}]%
        {Sharp2003RealTimeJava}
\bibfield{author}{\bibinfo{person}{D.~C. Sharp}, \bibinfo{person}{E. Pla},
  \bibinfo{person}{K.~R. Luecke}, {and} \bibinfo{person}{R.~J. Hassan}.}
  \bibinfo{year}{2003}\natexlab{}.
\newblock \showarticletitle{Evaluating real-time Java for mission-critical
  large-scale embedded systems}. In \bibinfo{booktitle}{{\em Proc. of the 9th
  IEEE Real-Time and Embedded Technology and Applications Symp.}}
  \bibinfo{pages}{30--36}.
\newblock


\bibitem[\protect\citeauthoryear{Vierhauser, Gr{\"u}nbacher, Egyed, Rabiser,
  and Heider}{Vierhauser et~al\mbox{.}}{2010}]%
        {vierhauserCC}
\bibfield{author}{\bibinfo{person}{Michael Vierhauser}, \bibinfo{person}{Paul
  Gr{\"u}nbacher}, \bibinfo{person}{Alexander Egyed}, \bibinfo{person}{Rick
  Rabiser}, {and} \bibinfo{person}{Wolfgang Heider}.}
  \bibinfo{year}{2010}\natexlab{}.
\newblock \showarticletitle{Flexible and scalable consistency checking on
  product line variability models}. In \bibinfo{booktitle}{{\em Proc. of the
  IEEE/ACM Int'l Conf. on Automated software engineering}}. ACM,
  \bibinfo{pages}{63--72}.
\newblock


\bibitem[\protect\citeauthoryear{Williams and Shin}{Williams and Shin}{2006}]%
        {iTrust}
\bibfield{author}{\bibinfo{person}{Laurie Williams} {and}
  \bibinfo{person}{Yonghee Shin}.} \bibinfo{year}{2006}\natexlab{}.
\newblock \showarticletitle{Work in progress: Exploring security and privacy
  concepts through the development and testing of the iTrust medical records
  system}. In \bibinfo{booktitle}{{\em Proc. of the 36th Annual Conf. on
  Frontiers in Education}}. IEEE, \bibinfo{pages}{30--31}.
\newblock


\end{thebibliography}

\end{document}